\documentclass[journal]{vgtc}                     % final (journal style)
\onlineid{0}

\usepackage{amsmath} 

\vgtccategory{Research}

\title{Passthrough Rigidity: \\The Behavioral and Visuomotor Costs of Mediated Perception}

\author{%
  \authororcid{Markus D.\ Solbach}{0009-0007-7288-9813},
  \authororcid{Mohit Goyal}{0000-0002-8835-158X},
  \authororcid{Sakar Khattar}{0009-0008-6707-959X},
  \authororcid{Jayant Varma}{0009-0001-9685-8807},\\
  \authororcid{Bjorn Vlaskamp}{0009-0007-5936-254X},
  \authororcid{John K. Tsotsos}{0000-0002-8621-9147},
  \authororcid{Konstantine Tsotsos}{0009-0008-2366-4666}
}

\authorfooter{
  \item
  	Markus D. Solbach is with York University.
  	E-mail: solbach@yorku.ca
  \item
  	Mohit Goyal is with Google, Inc.
  	E-mail: mohitgl@google.com
  \item
    Sakar Khattar is with Google, Inc.
    E-mail: sakark@google.com
  \item
    Jayant Varma is with York University
    E-mail: jayant@yorku.ca
  \item
    Bjorn Vlaskamp is with Google, Inc.
    E-mail: bjornvlaskamp@google.com
  \item
    John K. Tsotsos is with York University
    E-mail: tsotsos@yorku.ca
  \item
    Konstantine Tsotsos is with Google, Inc.
    E-mail: ktsotsos@google.com
}

\abstract{%
  Broad public adoption of head-mounted displays using video passthrough remains elusive despite significant market investment. A precise understanding of why users experience persistent discomfort even as hardware factors such as resolution and latency have dramatically improved remains an open issue. This paper investigates the impact of viewing the world through video passthrough systems on human behavioral and physiological patterns through a large-scale multimodal study. We developed a novel protocol to capture synchronized oculomotor, kinematic, and physiological data during a block assembly task requiring complex hand-eye coordination (Figure \ref{fig:illustration}). Using a within-subject design (N=110), we evaluated both natural and passthrough viewing conditions. Our results reveal a four-fold suppression of rotational head velocity and a pronounced decoupling of head-gaze coordination. This suggests motor caution being employed as an adaptive strategy – which we term ``Passthrough Rigidity''. This phenomenon appears to shift the information-gathering burden to the oculomotor system, resulting in significantly longer fixation durations and restricted visual search patterns. These kinematic shifts directly correlate with poorer task performance and measurable physiological cost, evidenced by a significant reduction in blink duration and increased reports of ocular strain and cognitive load. We conclude that current passthrough implementations induce a measurable shift from flexible exploration to motor caution, where task performance is preserved at the cost of user comfort and biomechanical efficiency. These findings provide a novel quantitative framework for evaluating and improving future XR devices, establishing that resolving ``comfort'' for passthrough requires addressing the deep-seated biomechanical compensations caused by mediated perception. 

  All data is made publicly available at \url{https://data.nvision.eecs.yorku.ca/Passthrough/}.
}


\teaser{
  \centering
  \includegraphics[width=.75\textwidth]{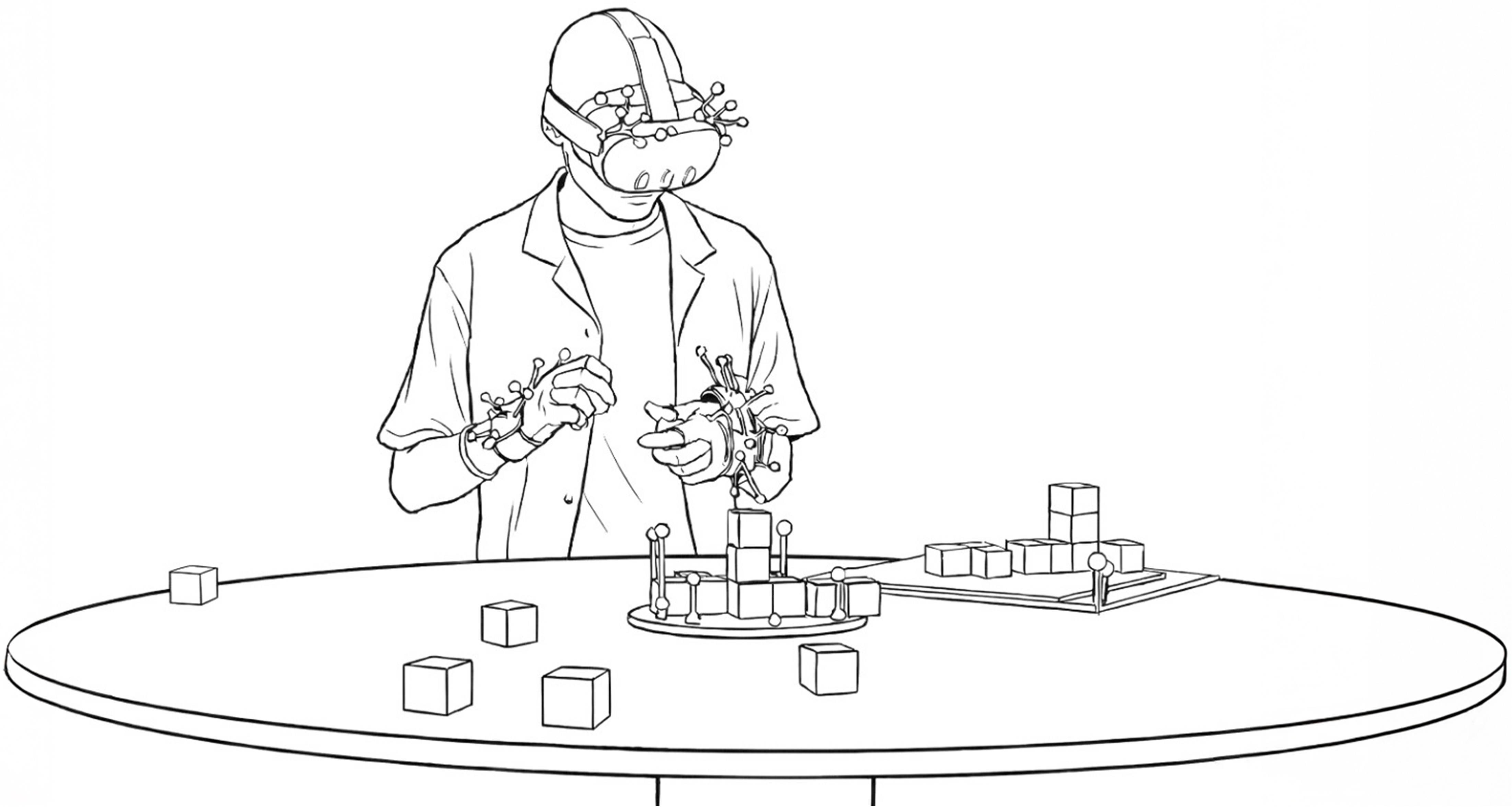}
    \caption{Illustration of a subject participating in our experiment. Equipped with head, hand, and eye-tracking devices, as well as passthrough system (Meta Quest 3).}
    \label{fig:illustration}
}

\graphicspath{{figs/}{figures/}{pictures/}{images/}{./}} % where to search for the images

\usepackage{booktabs}                  % only used for the table example
\usepackage{lipsum}                    % used to generate placeholder text
\usepackage{mwe}                       % used to generate placeholder figures
\usepackage{ccicons}                   % package to be able to use icons from creative commons

\usepackage{mathptmx}                  % use matching math font

\begin{document}

%%%%%%%%%%%%%%%%%%%%%%%%%%%%%%%%%%%%%%%%%%%%%%%%%%%%%%%%%%%%%%%%
%%%%%%%%%%%%%%%%%%%%%% START OF THE PAPER %%%%%%%%%%%%%%%%%%%%%%
%%%%%%%%%%%%%%%%%%%%%%%%%%%%%%%%%%%%%%%%%%%%%%%%%%%%%%%%%%%%%%%%

\firstsection{Introduction}
\label{sec:intro}
\maketitle

The landscape of human-computer interaction is undergoing a paradigm shift, moving from two-dimensional screens towards immersive, spatially-aware experiences. At the forefront of this evolution is Extended Reality (XR), an umbrella term encompassing Virtual Reality (VR), Augmented Reality (AR), and Mixed Reality (MR). While VR immerses users in entirely synthetic worlds and AR overlays digital information onto the real world, MR represents the most ambitious goal: the seamless blending of physical and virtual objects in a single, coherent environment where they can interact in real-time (Milgram et al. \cite{milgram1994taxonomy}). This pursuit has culminated in the recent growth of consumer and enterprise-grade MR headsets, such as the Samsung Galaxy XR, Apple Vision Pro and the Meta Quest 3, which promise to redefine domains ranging from surgical training and industrial manufacturing to remote collaboration and personal entertainment.

A critical enabling technology for modern MR is digital passthrough. Unlike optical see-through systems that use semi-transparent displays to superimpose graphics onto a user's direct view of the world, digital passthrough systems employ a different architecture. Forward-facing cameras on the headset capture a stereoscopic video feed of the user's surroundings, which is then computationally processed and displayed on opaque internal screens (Kress et al. \cite{kress2020optical}). This approach offers significant advantages, including the ability to achieve perfect occlusion (virtual objects can realistically block real-world objects), a wider field of view for digital content, and precise control over the color and brightness of the final composite image. Consequently, digital passthrough has become the dominant paradigm for achieving high-fidelity mixed reality.

However, despite its technological promise, a significant barrier to the wide\-spread adoption and prolonged use of passthrough MR is the prevalence of user-reported visual discomfort. As highlighted by Shibata et al. \cite{shibata2011zone} this phenomenon encompasses a range of adverse symptoms, including eye strain (asthenopia), fatigue, headaches, blurred vision, and in some cases, disorientation and nausea. While some of these issues, such as the vergence-accommodation conflict, are common to most stereoscopic displays, digital pass\-through introduces a unique and complex set of challenges. Unlike VR systems which immerse users in a fully virtual world where user movements are inherently decoupled from physical reality, video passthrough systems require users to coordinate physical actions within an optical reprojection of the real world. The final perceived image is not the real world itself, but a digital reconstruction, subject to a pipeline of potential deviations. These include motion-to-photon latency, image distortion from camera lenses, limited sensor resolution, digital noise in low-light conditions, and imperfect color reproduction (Maimone et al. \cite{6671761}). This discrepancy between the user's visual, vestibular, and proprioceptive systems can disrupt the brain's natural sensorimotor integration, leading to significant discomfort and a breakdown of presence.

A primary driver of this sensorimotor disconnect is the introduction of reprojection artifacts during the 3D reconstruction process. As highlighted by El Chemaly et al. \cite{el2025mind}, reprojection artifacts in mixed reality passthrough—often manifesting as visual warping or ``swimming'' effects—create spatial inaccuracies between the user's physical movements and their visual feedback. These geometric distortions significantly degrade fine motor task performance and hand-eye coordination, while simultaneously reducing overall user comfort by exacerbating visually-induced cybersickness. These specific visual artifacts are part of a broader taxonomy of perceptual issues inherent to AR and MR displays, which have been extensively studied over the years, ranging from foundational display limitations and perception problems (Kruijff et al. \cite{Kruijff2010PerceptualII}) to more recent, developer-oriented challenges in building modern video passthrough HMDs (Feld et al. \cite{10765260}).

As digital passthrough technology matures from a novelty into a foundational component of next-generation computing platforms, understanding and mitigating its associated visual discomfort is necessary. Failure to address these human factors will not only hinder user acceptance and limit application effectiveness but also pose potential risks to user well-being during extended use. Therefore, a systematic investigation into human comfort and task performance in passthrough systems is needed. 

The contribution of this research is two-fold. Firstly, we establish a novel methodology for recording multimodal data, capturing oculomotor, kinematic, and physiological data from subjects wearing passthrough devices, and design a block assembly task to mimic real-world visual challenges in a controlled setting. Secondly, we draw direct, evidence-based links between the physical and performance characteristics of a device and the resulting human visual behavior and performance. For example, we do not merely report that passthrough viewing increases subjective eye strain; we demonstrate how this reported discomfort is the experimental outcome of a shift in viewing behaviour. This shift reflects a degraded visual input, specifically the restriction of peripheral vision known in vision science as Tunnel Vision. Because this narrows the functional field of view, it prompts measurable changes in oculomotor strategy. Subjects intuitively retreat to a more limited exploratory behavior: they suppress large head translations and utilize shorter saccades with focused central fixations to compensate for the missing peripheral data. These changes further elicit a physiological irritation response (increased blink activity) as well as a slowing of hand movements. It is possible that these changes assist in minimizing latency and geometric re-projection artifacts, improve hand motion safety and control, and perhaps offer other compensatory benefits. Taken together, we term these effects ``Passthrough Rigidity''. By synthesizing these objective and subjective data streams, this study provides the first insights of the natural-passthrough visuomotor differences imposed by the current generation of passthrough technology, offering a more complete and nuanced picture of the human factors challenges that must be overcome to realize the vision of seamless spatial computing\footnote{Spatial Computing describes the creation of immersive and interactive experiences by augmenting the real environment with virtual elements in real time \cite{spatial}.}.

\section{Literature Review}
\label{sec:lit}

The advent of video passthrough head-mounted displays (HMDs) represents a significant step towards achieving mixed reality, where digital and physical worlds merge. This technology, which renders a real-time video feed of the user's environment inside an opaque headset, promises to enhance tasks ranging from industrial maintenance to surgical navigation. However, by mediating perception through an ``optical pipeline,'' passthrough systems introduce a host of human factors challenges that can affect performance and user comfort. A substantial body of research has emerged, i.e. Doughty et al. \cite{jimaging8070203} and Bailenson et al. \cite{bailenson2024seeing}, to investigate these challenges, typically by isolating specific technological limitations or their corresponding human responses. This section will synthesize the key findings from this literature to establish the context and identify the critical gap that the present study aims to fill.

\subsection{Cybersickness and Comfort in Video Passthrough}

The user experience in video passthrough HMDs has been colored by issues of cybersickness and visual comfort. Early investigations into this domain by Moss et al. \cite{moss2008simulator} established a strong link between viewing real-world scenes through an HMD and the onset of simulator sickness, demonstrating that symptoms tended to increase with the duration of exposure. This finding highlighted that the mere act of mediating vision, even of a familiar environment, was sufficient to induce discomfort.

Subsequent research has sought to identify the specific technological characteristics responsible for this effect. A primary factor is the sensory conflict that arises between the visual system and the vestibular system. Buker et al. \cite{buker2012effect} pinpointed system latency as a key driver of this conflict, showing that the temporal lag between a user's head movement and the corresponding update on the display is a significant cause of simulator sickness. Critically, their work also demonstrated that this is a tractable problem; by using predictive compensation algorithms to reduce the apparent latency, they were able to significantly reduce the magnitude of sickness experienced by users, directly linking a specific technological mitigation to an improved human-factors outcome.

Further work by Moss et al. \cite{moss2011characteristics} revealed that the scope of the user's view is also a critical factor. In their study, they found that simulator sickness was significantly greater when the user's peripheral vision was occluded by the HMD. This suggests that the absence of a stable, real-world visual reference frame exacerbates the sensory conflict, making users more susceptible to the negative effects of other display limitations like latency and image distortion.

Beyond motion-related sensory conflicts, a more fundamental challenge to visual comfort is the vergence-accommodation conflict (VAC), which is inherent to most stereoscopic displays. Hoffman et al. \cite{hoffman2008vergence} provided seminal work in this area, demonstrating that the requirement for the eyes to converge at the distance of a virtual object while accommodating (focusing) at the fixed distance of the HMD's screen is a primary source of visual fatigue and discomfort. Importantly, they showed that this conflict does not merely cause subjective discomfort; it directly hinders visual performance, reducing stereoacuity and increasing the time required to correctly identify stereoscopic stimuli. Together, this body of work establishes that cybersickness in video passthrough is a multi-faceted problem driven by a combination of technological limitations like latency and field-of-view occlusion, as well as fundamental physiological challenges like the vergence-accommodation conflict.

\subsection{Behavioral and Physiological Correlates of Video Passthrough Interaction}

To understand the mechanisms that translate technological limitations into the outcomes of cybersickness and altered task performance, researchers have examined the intermediate layer of behavioral and physiological adaptations. These objective metrics provide a window into the cognitive and physical costs of interacting with a mediated reality.

A primary behavioral adaptation is the shift in motor strategy to compensate for a restricted Field of View (FOV). As established by Moss et al. \cite{moss2011characteristics}, occluding the user's peripheral vision and the following change in visual scanning behavior is a significant driver of simulator sickness. With a limited FOV, users can no longer rely on efficient, low-effort eye movements (saccades) to survey their environment. Instead, they must engage in more frequent and larger-amplitude head movements, physically pointing the device's cameras to gather information that would normally be captured by the periphery. This shift from a natural ``eyes-first'' to a more cumbersome ``head-first'' search strategy represents a direct, measurable kinematic cost of using the device.

At a physiological level, the cognitive effort required to interpret a degraded and latent visual feed manifests in involuntary responses. One of the most powerful indicators of this cognitive load is the suppression of the natural blink rate. When concentrating on a digital screen, the human brain unconsciously reduces the frequency of blinking -- sometimes by as much as two-thirds -- to maximize the time available for information intake. This well-documented phenomenon is a primary contributor to digital eye strain and dry eye syndrome, as the reduced blinking disrupts the natural lubrication of the ocular surface (Kaur et al. \cite{kaur2022digital} and Kumari et al. \cite{kumari2024}. This physiological marker provides an objective correlate to the subjective feelings of eye strain and fatigue commonly reported by HMD users.

\subsection{Real-world trade-offs and Task Performance}

While the discussion of cybersickness focuses on internal physiological states, the utility of MR is determined by the user's ability to interact effectively with the external environment. Research into task performance in video passthrough indicates a functional trade-off: while MR systems provide spatially registered information, the optical mediation of the environment frequently impedes manual dexterity and fine motor control compared to unmediated reality.

Studies indicate that the mediation of the visual field imposes a measurable cost on interaction. Joyner et al. \cite{joyner2021comparison} investigated this phenomenon in the context of manual dexterity, observing that tasks requiring fine manipulation -- such as grasping small objects flush with a surface -- exhibit higher error rates and longer completion times in VR. They noted that users often failed to grasp target objects on the first attempt, a deficit attributed to the attenuation of precise depth cues and the occlusion of the user's fingertips by virtual elements. This suggests that spatial downsampling, lens distortions, and sensor noise in the video passthrough pipeline attenuate the fine, high-spatial-frequency visual cues (e.g., micro-textures and subtle shadows) required for seamless manual manipulation.

A primary factor contributing to this performance decrement is system latency and the resulting decoupling of visual and proprioceptive feedback. Unlike optical see-through systems where the real world is viewed directly, video passthrough systems introduce motion-to-photon latency for the entire environment. Rolland et al. \cite{rolland1995comparison} demonstrated that temporal mismatches between a user's proprioception (the felt position of the hand) and their visual perception (the seen position of the hand) interfere with the closed-loop control for precision tasks. Consequently, users tend to rely more heavily on visual feedback to correct their movements, resulting in trajectories that are less fluid and contain more corrective sub-movements, thereby increasing task duration.

Furthermore, the impact of these trade-offs extends beyond the immediate duration of use. Biocca et al. \cite{biocca1998virtual} established that these systems induce significant sensorimotor adaptation. Their research demonstrated that the spatial offset between the camera and the user's physical eye acts as a form of prism displacement, causing users to unconsciously recalibrate their hand-eye coordination. When users return to the unmediated real world, this recalibration persists as a negative aftereffect, leading to pointing errors and reduced coordination until the motor system readapts. This indicates that MR interaction can induce a temporary but measurable perturbation of the user's sensorimotor baseline.

\subsection{The Present Study: An Integrated, Multi-Modal Analysis}

While the existing literature has successfully identified and examined many of the individual challenges of video passthrough AR, these investigations have often been conducted in isolation. Studies typically focus on a single domain, such as the impact of FOV on head movement (Moss et al. \cite{moss2011characteristics}), the quantification of simulator sickness via questionnaires (Moss et al. \cite{moss2008simulator}), or the effect of AR on task errors (Moghaddam et al. \cite{moghaddam2021exploring}). One important aspect that is not well-studied is a holistic understanding of how these factors interrelate within a single, cohesive user experience.

The present study is unique in its integrated, multi-modal methodology. By simultaneously capturing oculomotor behavior (saccades, fixations), physiological indicators (blink rate), gross motor kinematics (head and hand movement), and detailed subjective reports (SSQ), this work moves beyond identifying isolated deficits to characterizing the user's adaptive strategy.

Based on the existing literature, this study compares user behavior in natural versus video passthrough viewing conditions to test the following hypotheses:

\begin{itemize}
\item[H1] Kinematic factors: Kinematic velocities will be significantly lower in the video passthrough condition compared to unmediated natural vision. Specifically, we hypothesize a substantial decrease in rotational head velocity, alongside measurable reductions in the translational and rotational velocities of the hands.  
\item[H2] Oculomotor factors: To compensate for lower kinematic velocities we expect occulomotor system to take the burden of spatial exploration. Therefore, we hypothesize that video passthrough viewing will result in longer fixation durations, longer saccade durations, and shorter blink durations compared to natural viewing.  
\item[H3] Performance \& Comfort: The video passthrough viewing condition will negatively impact both objective task outcomes and subjective user experience. We hypothesize that participants will exhibit lower manual task accuracy and longer response times, while reporting higher levels of visual discomfort, cognitive load, and overall fatigue (measured with self-reported SSQ).  
\end{itemize}

\section{Methods}
\label{sec:meth}
% This Section is good to go!

To investigate the sensory-motor and cognitive impacts of mediated perception, we designed a multi-modal experiment to directly compare human performance during a complex hand-eye coordination task under two distinct viewing conditions: unaided natural vision and video passthrough augmented reality (using Meta's Quest 3). We employed a within-subjects design to control for individual differences. For each subject, we recorded a dataset including kinematic tracking of the head and hands, detailed oculomotor tracking, first- and third-person video, and a comfort questionnaire after each trial. The following sections provide details of participants, experimental design, procedures, apparatus, and data collection.

\subsection{Task and Experimental Design}

The study employed a within-subjects design to investigate the effects of mediated vision on human performance.

The task was a block assembly challenge, requiring significant hand-eye coordination (e.g. Hayhoe et al. \cite{hayhoe2005eye}, Melnik et al. \cite{melnik2018world} and Ma et al. \cite{ma2025phyblock}). Similar past studies used static tasks within limited depth planes. To involve greater depth of field activity, 3D hand-eye coordination, and trigger multiple views, we introduced a motorized turntable on which the task was performed. Specifically, this dynamic apparatus was introduced to compel participants to continuously update their 3D visuospatial problem-solving models, actively preventing static, in-hand assembly. For each trial, participants were instructed: ``Using the parts on the table, assemble the object you see in the center of the table.'' The task was governed by a set of rules to ensure consistency: participants began at a fixed starting position, built their assembly in a fixed area (assembly area) on the turntable, and were explicitly instructed not to perform in-hand assembly or pile up parts. It is important to note that our study focuses exclusively on the visual reprojection of the physical environment; all building blocks and target objects were purely physical, and no virtual or digital content was embedded in the scene.

The primary independent variable was the viewing type, with two levels: \textit{Natural\footnote{With \textit{Natural} we describe the viewing condition in which a participant wears an eye tracking instrument. While the instrument is lightweight, untethered and does not obstruct the participant's view, we have not investigated the impact the instrument might have.}} and \textit{Passthrough}. The experiment also manipulated three other variables to assess performance across a range of conditions. 

\begin{figure}
\centering
        \includegraphics[width=1.0\linewidth]{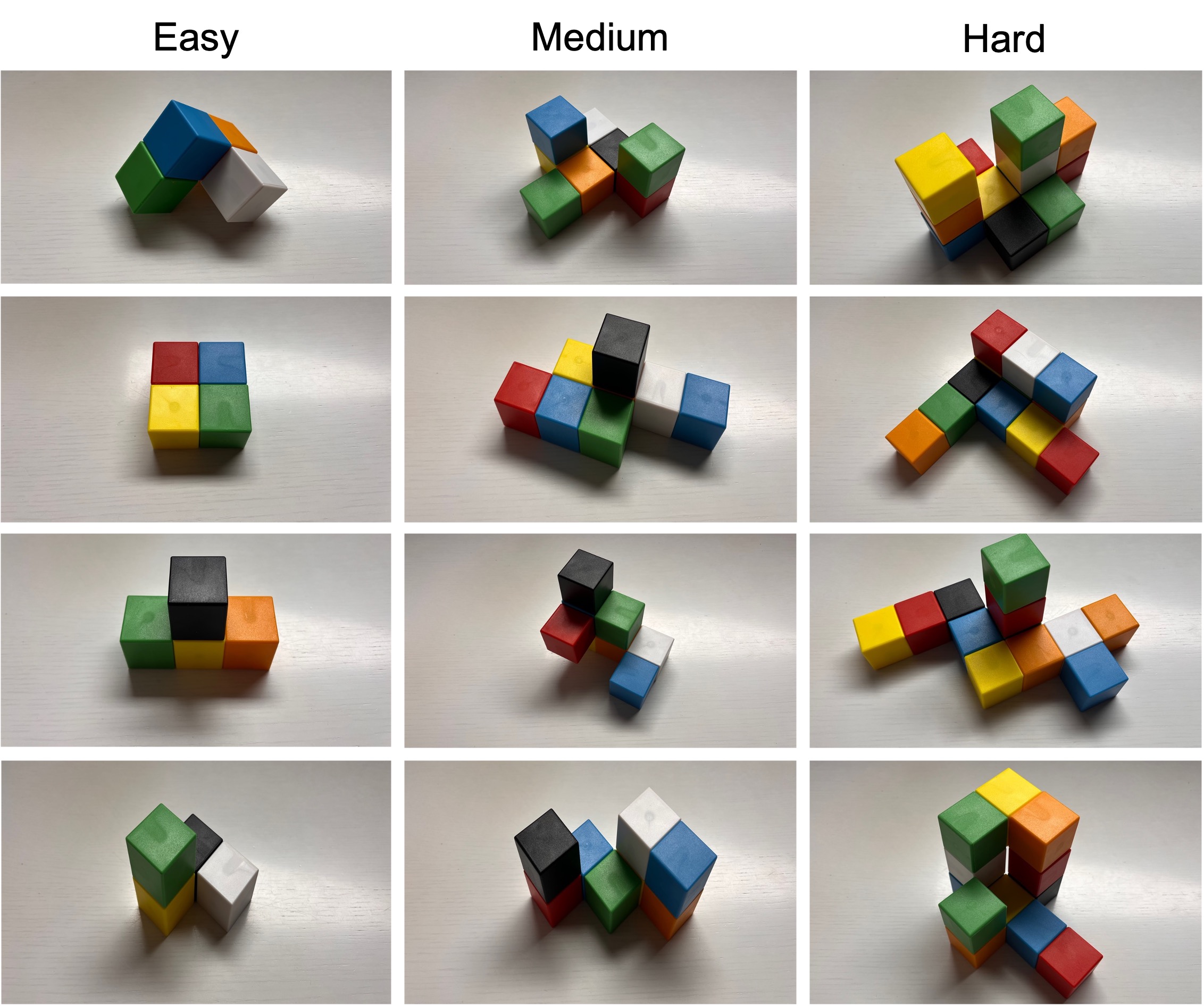}
    \caption{Stimuli used in this study. 3 different complexity levels with 4 variations each.}
    \label{fig:stimuli}
\end{figure}

Cuboids connect magnetically into an aligned position automatically. Target objects were of three levels of complexity, i.e., constructed with 4, 8 or 12 cuboids (Easy, Medium, Hard). A cuboid is $2.5cm \times 2.5cm \times 2.5cm$ in size and is colored in one of the following 8 colors: white, black, yellow, orange, red, green, blue, and purple. Four unique cuboid constructions (objects) were available for each complexity level to prevent repetition across trials. The configurational complexity was kept similar across all four variations. Figure \ref{fig:stimuli} illustrates the objects used in the experiment. 

Additionally, we explored four different rotational speed settings of the turntable; 0, 5, 10 and 15 RPM. Lastly, in selected trials some subjects were permitted to move while others were not in order to determine the differences. ``Subject Movement'' has two levels; In the ``Stationary'' condition, participants were required to remain at a fixed starting position and in the ``Moving'' condition, they were allowed to move freely around the apparatus. The 0 RPM turntable speed was only used in combination with the Moving condition.

This brings the entire configuration space to 

\begin{equation}
\begin{aligned}
CS = \underbrace{\textrm{Speed} (4) \times \textrm{Complexity} (3) \times \textrm{Viewing Condition} (2)}_\textrm{free movement} \\+ \underbrace{\textrm{Speed} (3) \times \textrm{Complexity} (3) \times \textrm{Viewing Condition} (2)}_\textrm{stationary} = 42. 
\end{aligned}
\end{equation}

\subsection{Participants}

A total of 110 participants (M = 24.6 years of age, SD = 3.8 years) were recruited for this study. The recruitment was done through public postings through blackboards, flyers, and group emails at York University. Given this recruitment strategy, the participant profile consisted primarily of university students (84\%), with the remaining 16\% drawn from the broader general public. Participants reported varying levels of prior Extended Reality (XR) experience: the majority (51\%) had no prior experience, 31.5\% rarely used an XR device, 13.4\% reported occasionally, 2.5\% daily and  1.6\% frequent usage. Gender identification was not recorded for this study. The experiment was granted ethics approval by the office of research ethics at York University (Certificate \#e2024-248).

In this study, color-perception and stereopsis is critical for comparing natural and passthrough-mediated vision. All participants were screened for color-blindness using standard  Ishihara color plates prior to the stereoacuity check. We used the RANDOT Stereotest kit to measure stereoacuity in arc seconds. To minimize variability from stereo-blindness or significantly reduced stereo vision, which as pointed out by Dorman et al. \cite{dorman201750}, affects 5-30\% of the population, we sought to recruit participants with normal stereo acuity, defined as 20 arc seconds or better. The results of each participant's stereopsis test were recorded as part of their demographic data.

\subsection{Procedure}

Each participant attended a single session lasting approximately one hour. Upon arrival and after providing informed consent, they underwent the RANDOT stereopsis screening, color-blindness test, and vision test. Participants were then fitted with the appropriate headgear and hand-tracking gloves.

Each session comprised 16 trials, split evenly between counterbalanced Natural and Passthrough viewing conditions to mitigate order effects. Within each 8-trial block, independent variables (object complexity, turntable speed, and allowed movement) were strictly counterbalanced using a Latin square design. After each trial, participants completed the Simulator Sickness Questionnaire (SSQ) by Kennedy et al. \cite{kennedy1993simulator}) to assess their subjective experience. The questionnaire asks participants to rate their discomfort in 16 categories on a scale of four different levels; none, slight, moderate and severe. The 16 different categories are as follows: General Discomfort, Fatigue, Headache, Eye Strain, Difficulty Focusing, Increased Salivation, Sweating, Nausea, Difficulty Concentrating, Fullness of the Head, Blurred Vision, Dizzy (eyes open), Dizzy (eyes closed), Vertigo (Giddiness), Stomach Awareness, and Burping. For the final analysis, standard Kennedy et al. \cite{kennedy1993simulator} weights were applied to absolute change scores (deltas from baseline) to quantify the physiological impact.

%At the conclusion of the entire session, a final post-experiment questionnaire was administered to gather overall feedback.

\subsection{Apparatus and Materials}

\begin{figure*}[h!]
\centering
        \includegraphics[width=1.0\textwidth]{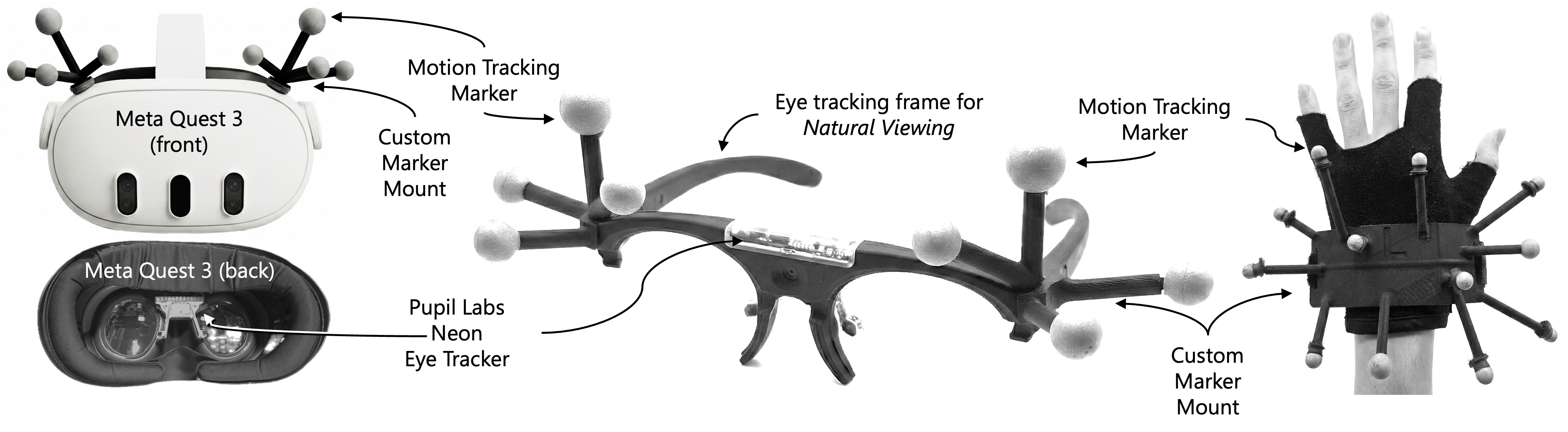}
    \caption{Meta Quest 3 HMD with custom-designed motion tracking attachments (left), Pupil Neon eye-tracking glasses (without lenses) with our custom-designed motion tracking attachments (middle) and hand-tracking gloves with custom-designed motion tracking attachments (right).}
    \label{fig:tracking}
\end{figure*}

The experiment was conducted using a custom-integrated data acquisition platform, PESAO by Solbach et al. \cite{solbach2021tracking}, which is specifically designed to record detailed data of active observers in the real-world using a combination of synchronized motion tracking, eye tracking and camera systems. This platform has been central to recent studies of human real-world 3D visuospatial problem-solving (Solbach et al. \cite{solbach2023psychophysics}) and visual search behaviors (Wu et al. \cite{wu2025real}.

\begin{figure}[ht!]
\centering
        \includegraphics[width=1.0\linewidth]{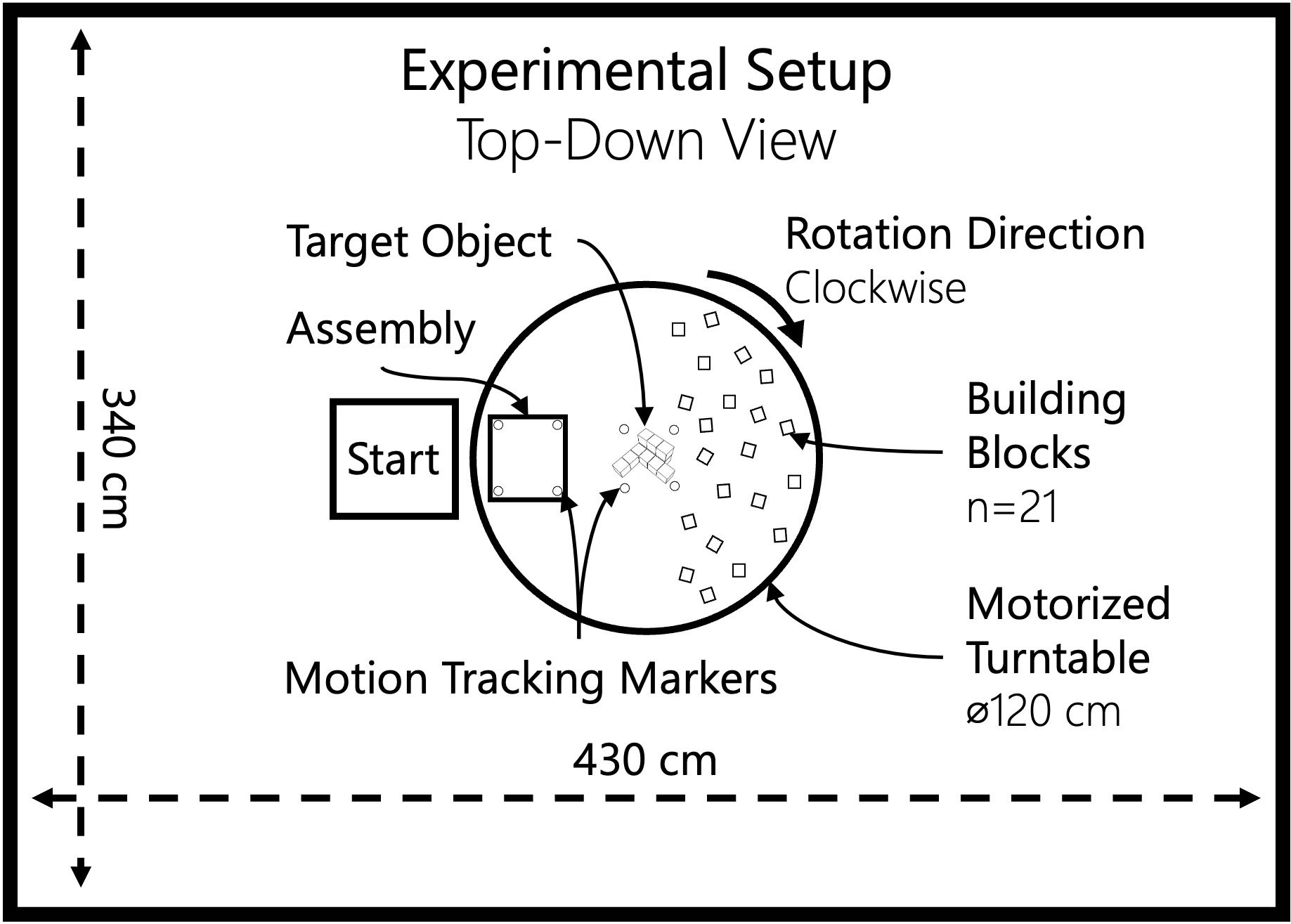}
    \caption{Top-Down View of the Experimental Setup.}
    \label{fig:experimental}
\end{figure}

Figure \ref{fig:experimental} illustrates a top-down view of the experimental setup. The core apparatus consisted of a central turntable for presenting stimuli. The turntable is 120cm in diameter and is motorized to spin at different speeds clockwise and height can be adjusted to suit the subject. The turntable is tracked using motion markers and the target object is placed in its center. A dedicated assembly area is marked on one side of the table to indicate where the object has to be built, building blocks are placed across from it on the other side of the table. The entire tracking volume and space in which the participant can move in measures $430cm \times 340cm$.

The participant's head and hand movements were tracked using a high-fidelity motion tracking system (OptiTrack) at 120 Hz. For the hand-eye coordination task, participants wore tracking gloves equipped with markers to capture the 6D pose of each hand. The eyes were tracked using the mobile eye tracker from Pupil Neon \cite{neon}. The only passthrough device available for this first set of experiments is the Meta Quest 3 specifically due the immediate availability of Pupil attachments. We equipped the Meta Quest 3 HMD with custom tracking markers as well. Figure \ref{fig:tracking} shows the participant-worn tracking setup, except for the smartphone that is used to run the eye-tracker. The smartphone is wired to the eye-tracker via an USB cable and carried in pocket of the subject.

\begin{figure}[h!]
\centering
        \includegraphics[width=1.0\linewidth]{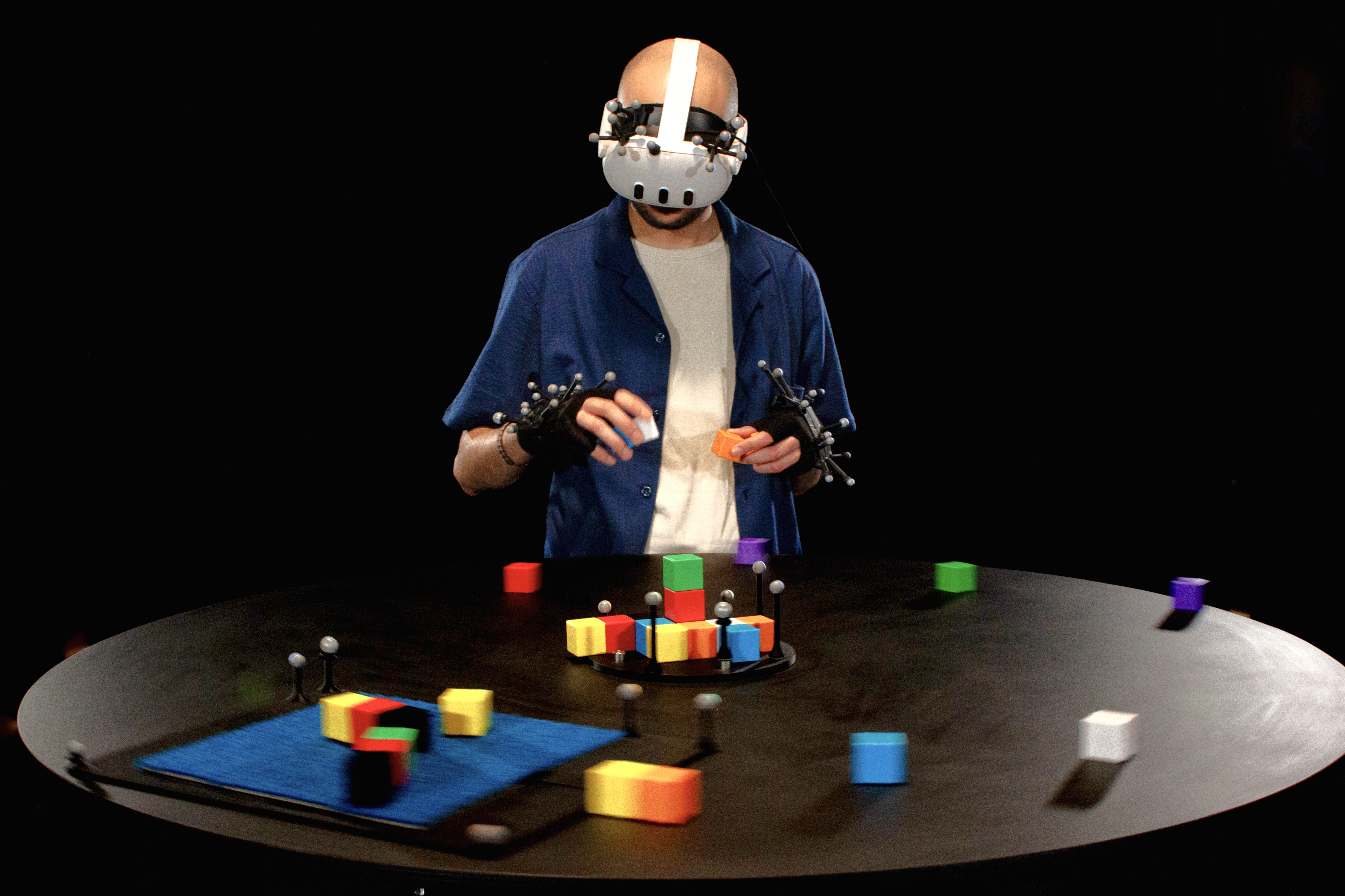}
    \caption{A participant doing a trial in passthrough viewing. In this example, the table was spinning at 10 RPM, the subject had to remain stationary, the presented object to copy was of Hard complexity.}
    \label{fig:exp_photo}
\end{figure}

The experiment featured two distinct viewing conditions:

\begin{itemize}
\item Natural Viewing: Participants wore Pupil Neon eye-tracking glasses, which recorded binocular eye data at 200 Hz and a first-person scene camera view ($1600px \times 1200px$ at 30 Hz).

\item Passthrough Viewing: Participants wore a Meta Quest 3 HMD. This specific device was chosen because it seamlessly accommodated the Pupil Neon eye-tracking system via an off-the-shelf mount without adding significant weight to the user. Other devices like the Apple Vision Pro, with built-in eye tracking, do not provide access to the eye-tracking data and thus need external attachments which were not available to us. Prior to testing, the device's hardware IPD wheel was explicitly tailored to each participant's measured inter-pupillary distance. The device's video see-through functionality provided the view of the environment, and the left eye's passthrough video feed was recorded ($1032px\times2064px$ at 60 Hz). The device was equipped with a special frame to enable installing the Pupil Neon eye-tracking system. Figure \ref{fig:exp_photo} shows a photo of a participant doing a trial in the passthrough viewing setting. 
\end{itemize}

To ensure the passthrough system operated under optimal conditions, ambient and task-area lighting was standardized and monitored. A Yoctopuce lux meter was used to measure illuminance once per participant. Measurements were taken on the surface of the assembly table and at three locations on the floor around the turntable (averaged for an ambient reading) to confirm light levels exceeded Quest 3's recommended minimum of 50 lux.

\subsection{Data Collection and Measures}

The PESAO platform was used to record a comprehensive, multi-modal dataset with 6 synchronized data streams as described here:

\textbf{Kinematic Data (120 Hz):} The 6D pose (position and orientation) of the participant's head (via the HMD or eye tracker), both hands, the turntable, and the designated assembly area were recorded.

\textbf{Oculomotor Data (200 Hz):} A Pupil Neon eye-tracker is used to capture 2D and 3D gaze data, along with classified eye events including fixations, saccades, and blinks. 

\textbf{Video Data:} Three video streams were captured: a first-person scene camera from the Pupil Neon headset (30 Hz, Natural only), the left-eye passthrough video from the Quest 3 (60 Hz, Passthrough only), and a third-person ``bird's-eye'' view of the experimental setup (1$920\times1080px$ at 30 Hz).

\textbf{IMU Data (110 Hz):} 9-DoF inertial measurement unit data was recorded from the Pupil Neon device.

\textbf{Subjective Data:} The 16-item SSQ was administered after each trial. 

%A paper-based post-experiment questionnaire was completed once per participant.

\textbf{Control and Environmental Data:} The settings for all independent variables were logged for each trial. Subject demographics and illuminance measurements were logged once per experiment.

\textbf{Sampling of Configuration Space:} 

Our entire configuration space involves 42 configurations. Because we recruited 110 subjects who each completed only 16 trials, the experiment utilized a balanced, fractionally sampled incomplete block design. To ensure robust data collection, the specific condition configurations were pseudorandomly assigned across the 110 participants. This strategy brings our total number of samples for the entire configuration space to approximately 41.9 per condition ($\frac{110\times16}{42}$). Our sampling rate aligns with established psychophysical standards for minimizing parameter estimation error by Prins et al. \cite{prins2016psychophysics} and Garc{\'\i}a-P{\'e}rez \cite{garcia2005sampling}.

\subsection{Statistical Analysis}
\label{sec:stat_analysis}

Normality and homoscedasticity assumptions were evaluated prior to conducting repeated-measures ANOVAs. As anticipated for large samples ($N > 1300$), Shapiro-Wilk tests indicated deviations from normality ($p < .05$). Levene's tests confirmed equal variances for translational kinematics and vestibular symptoms ($p > .05$), but revealed unequal variances for rotational kinematics and ocular strain ($p < .05$) -- reflecting the specific behavioral constraints of passthrough viewing. Given ANOVA's established robustness to these violations in large datasets, parametric models were retained. We report partial eta squared ($\eta_p^2$) to evaluate practical significance and mitigate $p$-value over-sensitivity.

\section{Results}
\label{sec:res}
% This Section needs work
%% New 'comparative' histograms. 
%% Adding significant analysis.

The recorded empirical data, encompassing oculomotor behavior, head and hand kinematics, and physiological indicators reveal a significant and multi-faceted set of differences associated with the use of the passthrough HMD. Primarily, they show a degradation of isolated skills; however, taken together they reveal a significant adaptation in the user's sensory-motor strategy. To navigate and interact with the world through a technologically mediated interface, users adopt a series of consistent, measurable, and resource-intensive compensatory behaviors.

In the following sections we will present results organized by performance (accuracy and response time), kinematic (Head and Hand Movements), oculomotor (Gaze and Eye Movements) and lastly, physiological (Simulator Sickness Questionnaire) measures. We will also report normative values for comparison from the literature where available. All plots in this section illustrate the mean (circle) and 95\% confidence interval (whiskers). 

%The principal findings indicate that the passthrough condition forces a transition from an efficient, eye-dominant visual search strategy to a more cumbersome and metabolically costly head-dominant strategy. This adaptation is a direct response to the technological limitations of the HMD, most notably its restricted field of view (FOV). Furthermore, the analysis uncovers profound evidence of increased cognitive load, physiologically manifested as a severe suppression of the natural blink rate. This finding suggests that the very act of perceiving and interpreting the passthrough video feed is a cognitively demanding task, imposing a baseline perceptual load that precedes any task-specific cognitive effort. This heightened load has significant implications for user fatigue, comfort, and the long-term viability of passthrough technology for extended use, potentially impacting ocular health. These objective behavioral shifts are strongly corroborated by subjective user reports, with the passthrough condition inducing significantly higher rates of simulator sickness, including eye strain, general discomfort, and difficulty concentrating. While gross motor control of the hands appears resilient for the task studied, this may mask underlying deficits in depth perception and fine motor control that are critical for safety and performance in more complex scenarios. In conclusion, the data shows that the human sensory-motor system working demonstrably harder and less efficiently to overcome the inherent constraints of current-generation passthrough technology.

\subsection{Analysis of Performance}

We first examined the effects of the viewing condition on task efficiency and success. The analysis focused on two primary metrics: Response Time (seconds) and Correctness (accuracy percentage).

\begin{figure}[h!t]
\centering
\begin{minipage}[t]{.5\linewidth}
  \centering
    \includegraphics[height=5cm]{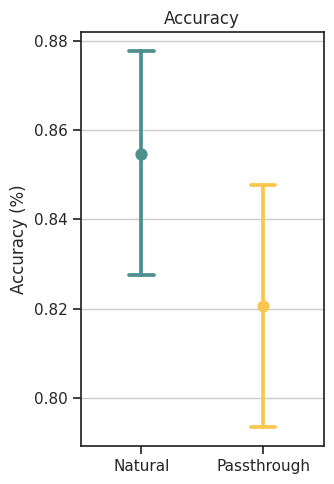}
    \captionsetup{margin=5pt}
    \caption{Accuracy (\%) of successful trials. Shown is mean and 95\% confidence interval for natural (left) and passthrough (right) viewing.}
    \label{fig:acc}
\end{minipage}%
\begin{minipage}[t]{.5\linewidth}
  \centering
  \includegraphics[height=5cm]{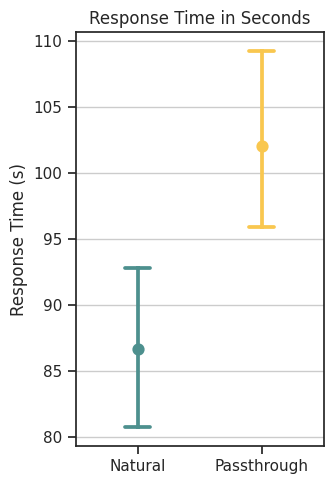}
  \captionsetup{margin=5pt}
  \caption{Response Time (s) required to complete a trial. Measurement spans from the start signal to the participant's voluntary termination, indicating they were either satisfied with the assembly or chose to end the trial.}
    \label{fig:response_time}
\end{minipage}
\end{figure}

\textbf{Accuracy} A repeated-measures ANOVA revealed a significant but practically trivial effect of Viewing condition on Accuracy ($F_{1, 1329}=4.07$, $p=.044$, $\eta_p^2 = .00$)\footnote{Following standard APA statistical formatting, ANOVA results report the $F$-statistic (the ratio of between-group to within-group variance) and degrees of freedom in addition to the $p$-value (Field \cite{field2024discovering}).}. Participants achieved a higher mean accuracy using Natural vision (M = $85.46$\%, SD = 35.3\%) compared to the Passthrough system (M = $82.07$\%, SD = 38.4 \%). Figure \ref{fig:acc} illustrates the accuracy for natural and passthrough settings across all trials. 

%Figure \ref{fig:acc} depicts the fluctuation of accuracy across the eight experimental trials. While the Natural condition (teal line) generally sustains higher accuracy, the performance gap varies across specific trials. Notably, while the ANOVA indicated no significant main effect for the Trial variable on Correctness, the visual data in Figure \ref{fig:acc} shows distinct peaks and dips, including a convergence of performance scores in the final trial (Trial 8), where Passthrough accuracy momentarily exceeded Natural accuracy.

\textbf{Response Time} The Viewing condition also yielded a significant main effect on Response Time with a small effect size ($F_{1, 1379}=14.2$, $p<0.05$). Participants completed the assembly tasks significantly faster in the Natural viewing condition (M = $86.62$ s, SD = 8.73 s) compared to the Passthrough condition (M = $102.03$ s, SD = 9.61 s). Figure \ref{fig:response_time} illustrates response times over all trials for natural and passthrough.

\subsection{Analysis of Oculomotor Behavior}

We further analyzed the impact of the viewing condition on oculomotor behavior, specifically examining fixation dynamics, saccade metrics, and blink duration.

\textbf{Fixation Dynamics} The viewing condition had a significant main effect on both the temporal and frequency domains of fixations. For Fixation Duration, participants exhibited significantly longer fixations in the Passthrough condition (M = $304.14$ ms, SD = 79.79 ms) compared to the Natural condition (M = $279.26$ ms, SD = 55 ms), $F_{1, 1379}=50.3$, $p<0.001$, $\eta_p^2 = .04$. 

Conversely, Fixation Frequency showed the inverse pattern with a moderate effect size: the Natural condition has a higher frequency of fixations (M = $2.71$ Hz, SD = 0.32 Hz) compared to the Passthrough condition (M = $2.46$ Hz, SD = 0.41 Hz), $F_{1, 1379}=179$, $p<0.001$, $\eta_p^2 = .11$. Natural and Passthrough settings fell within the typical range of Fixation Duration for scene viewing (260-330 ms, Rayner \cite{rayner1998eye} and Land et al. \cite{land2001ways}, but exhibit a slower frequency (typical 3-4 Hz, Findlay et al. \cite{findlay2003active}).

\begin{figure}[h!t]
\centering
\begin{minipage}[t]{.5\linewidth}
  \centering
  \captionsetup{margin=5pt}
  \includegraphics[height=5cm]{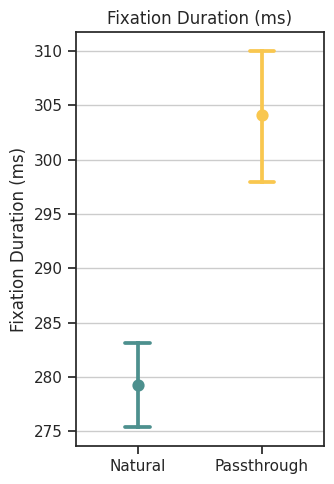}
  \captionof{figure}{Fixation Duration (ms) over all $\approx1,440$ trials. About 250,000 fixations in total.}
  \label{fig:fix_dur}
\end{minipage}%
\begin{minipage}[t]{.5\linewidth}
  \centering
  \captionsetup{margin=5pt}
  \includegraphics[height=5cm]{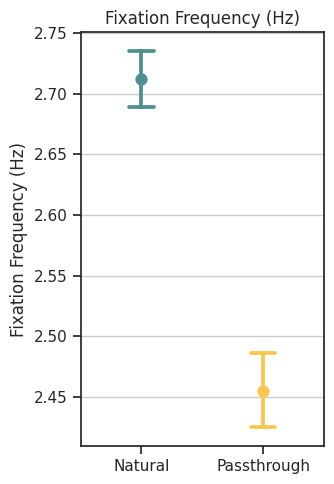}
  \captionof{figure}{Frequency (Hz) over all trials. Calculated as number of fixations divided by response time.}
  \label{fig:fix_freq}
\end{minipage}
\end{figure}

The boxplots in Figure \ref{fig:fix_dur} and \ref{fig:fix_freq} visually confirm these shifts. The distribution for Fixation Duration (Figure \ref{fig:fix_dur}) shows a higher mean and a more dispersed confidence intervals Passthrough condition. In contrast, the Fixation Frequency plot (Figure \ref{fig:fix_freq}) illustrates a reduction in the density of high-frequency fixation events when using the passthrough system.

%The data for Fixation Duration and Fixation Frequency confirm these shifts. Fixation Duration has a higher mean a more dispersed standard deviation in Passthrough condition: Natural (M=279.26ms, SD=55ms) and Passthrough (M=304.14 ms, SD=79.79 ms). In contrast, the Fixation Frequency exhibits a reduction in the standard deviation of high-frequency fixation when using the passthrough system: Natural (M=271 Hz, SD=0.32Hz) and Passthrough (M=2.46 Hz, SD=0.18 Hz)

\textbf{Spatial Distribution of Fixations} Beyond temporal dynamics, the analysis revealed significant differences in the spatial distribution of fixation locations.

\begin{figure}[h!t]
\centering
        \includegraphics[width=1.0\linewidth]{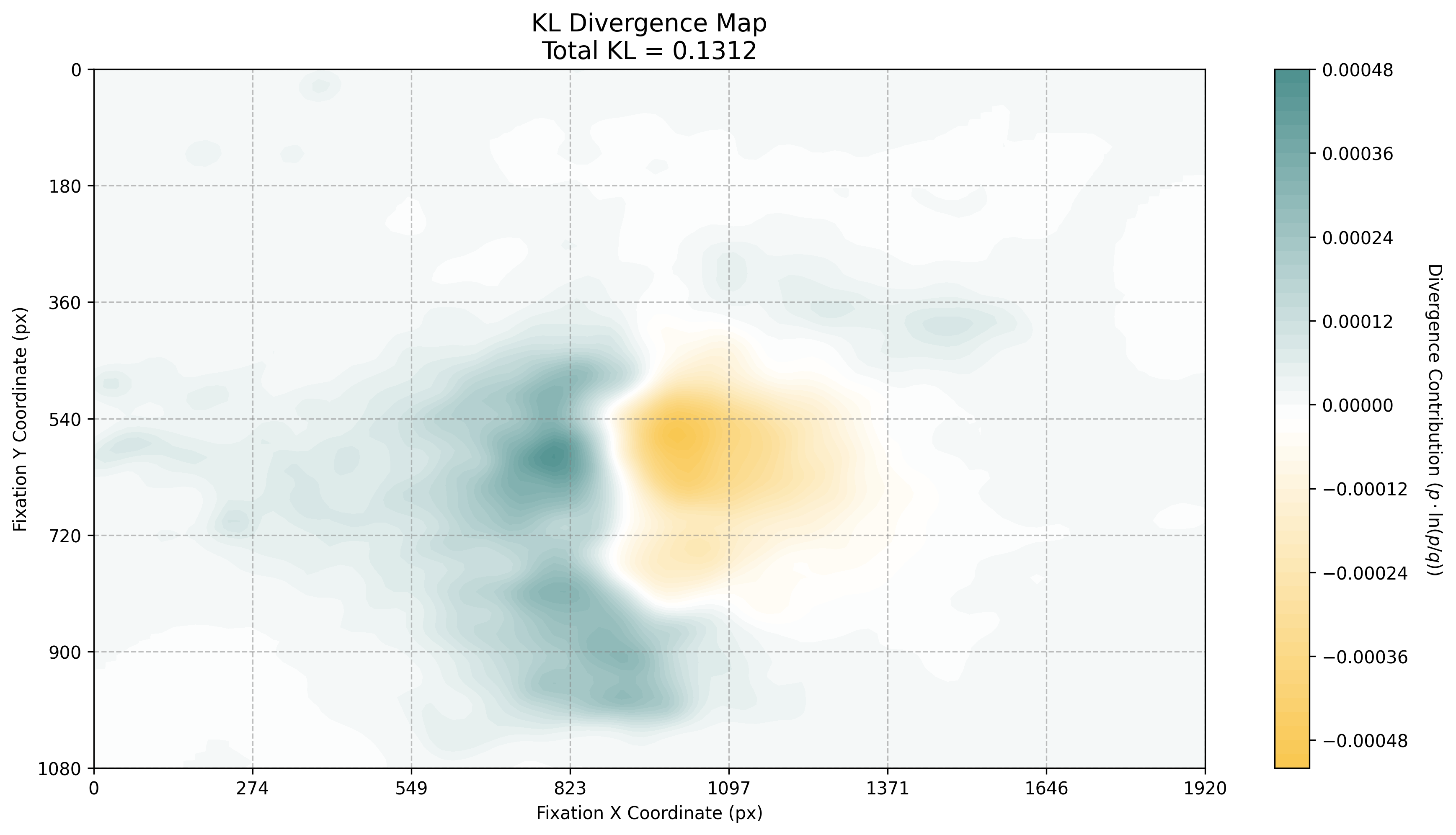}
    \caption{Kullback–Leibler divergence Map of passthrough and natural fixational distributions. The yellow colored center depicts that passthrough fixations are more centrally focused.}
    \label{fig:fix_kl}
\end{figure}

Figure \ref{fig:fix_kl} illustrates the Kullback–Leibler divergence (Kullback et al. \cite{kullback1951information}) map of passthrough and natural fixational distributions with $KL = 0.1318$. It is clearly visible that the passthrough fixations are more centrally focused (yellow colored center) and natural fixations, in comparison, are wider spread (teal). A $KL$ value of $0.1318$ signifies a moderate, measurable difference between the two distributions and therefore, implies a change in eye movement behaviour (Bylinskii et al. \cite{bylinskii2018different}). Given that the goal of passthrough devices is to be as faithful to natural vision as possible, this degree of divergence is significant.

\textbf{Saccade Metrics} Analysis of saccadic behavior revealed significant differences in both amplitude and duration. Here we measure the duration from the onset of the saccade (leaving current fixation) to the stabilization of the gaze (arriving at next fixation), the amplitude is measured in degree of visual angle (distance travelled). There was a significant but small main effect of viewing type on Saccade Amplitude, with participants performing larger saccades in the Natural condition (M = $12.53^\circ$, SD = $2.65^\circ$) compared to the Passthrough condition (M = $11.90^\circ$, SD = $2.56^\circ$), $F_{1, 1379}=30.6$, $p<0.001$, $\eta_p^2 = .02$.

However, the most pronounced effect was observed in Saccade Duration. Saccades were significantly longer in the Passthrough condition (M = $134.61$ ms, SD = $35.49$) compared to the Natural condition (M = $92.68$ ms, SD = $31.68$ ms), representing a statistically significant but trivial effect, $F_{1, 1379}=9.92$, $p=0.002$, $\eta_p^2 = .01$.

%\begin{figure}[h!t]
%\centering
%\begin{minipage}[t]{.5\linewidth}
%  \centering
% \captionsetup{margin=5pt}
%  \includegraphics[height=6cm]{img/selection/Occulomotor/bar_Saccade_Duration_(ms).png}
%  \captionof{figure}{Saccade Duration in ms.}
%  \label{fig:sacc_dur}
%\end{minipage}%
%\begin{minipage}[t]{.5\linewidth}
%  \centering
%  \captionsetup{margin=5pt}
%  \includegraphics[height=6cm]{img/selection/Occulomotor/bar_Saccade_Amplitude_(degree).png}
%  \captionof{figure}{Saccade Amplitude in degree of visual angle.}
%  \label{fig:sacc_amp}
%\end{minipage}
%\end{figure}

%The plot for Saccade Duration (Figure \ref{fig:sacc_dur}) displays a dramatic separation between conditions, with the range for Passthrough residing almost entirely above the median of the Natural condition. The Saccade Amplitude plot (Figure \ref{fig:sacc_amp}) shows a subtler compression of the range in the Passthrough condition. 

Compared to normative behaviour, Saccade Amplitudes for either mode are within the range of up to $130^\circ$ for an active task (Land et al. \cite{land1999roles}). However, Saccade Durations for Passthrough are well beyond the 80-100 ms range (Land et al. \cite{land1999roles}).

\textbf{Blink Duration} Finally, we examined Blink Duration as a metric of ocular engagement. The duration is measured from the start of eye-lid closure to the end of reopning. The ANOVA revealed a significant main effect of viewing condition, $F_{1, 1379}=217$, $p=0.002$, $\eta_p^2 = .14$. Contrary to the fixation and saccade duration trends, blink durations were significantly shorter in the Passthrough condition (M = $228.64$ ms, SD = $42.11$ ms) compared to the Natural condition (M = $261.42$, SD = $40.58$ ms). For either viewing type, the Blink Duration is within the normative reported values of 100-400 ms (Stern et al. \cite{stern1984endogenous}).

%The distribution shown in Figure \ref{fig:blink_dur} reflects this decrease, with the Passthrough condition exhibiting a lower median and a tighter clustering of data points at the lower end of the time scale compared to the broader distribution observed in Natural viewing. 

%\begin{figure}[h!]
%\centering
%        \includegraphics[width=0.5\linewidth]{img/selection/Occulomotor/bar_Blink_Duration_(ms).png}
%    \caption{Blink Duration in ms. Measured from complete eye lid closing to start of opening.}
%  \label{fig:blink_dur}
%\end{figure}

\begin{figure}[h!]
\centering
        \includegraphics[width=1.0\linewidth]{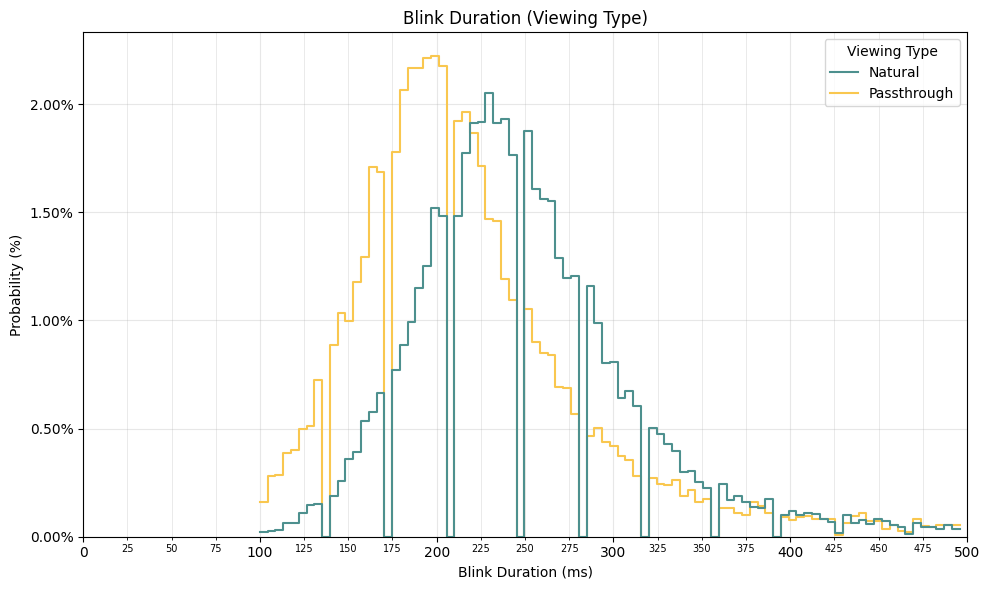}
    \caption{Histogram of Blink Duration. The visible gaps in the histogram are representative of the blink sampling frequency of the neon pupil eye tracker; every 45ms a 5ms section is binned into the previous and next section.}
  \label{fig:blink_hist}
\end{figure}

Figure \ref{fig:blink_hist} presents the histogram of blink durations for both viewing types. While the overall shape of the distribution remains similar across conditions—characterized by a positive skew typical of duration metrics—there is a clear leftward shift in the Passthrough distribution (yellow), corresponding to the shorter mean duration. Additionally, the Passthrough distribution appears somewhat narrower compared to the Natural condition (teal), suggesting less variability in blink duration when using the mediated vision system. The Blink Frequency, however, was not significantly affected by the viewing type; natural and passthrough viewing resulted in 0.25 Hz and 0.24 Hz, respectively.

\subsection{Analysis of Kinematics}

We examined the kinematic profiles of both the head and hands to understand how mediated vision influenced motor strategies. This analysis encompasses translational velocities (m/s), rotational velocities (degree/s), and the coordination of head rotation and visual fixations.

\textbf{Head Kinematics} The viewing condition significantly altered head movement dynamics. For translational movement, participants moved their heads significantly faster in the Natural condition (M = $0.21$ m/s, SD = $0.098$ m/s) compared to the Passthrough condition (M = $0.18$ m/s, SD = $0.093$ m/s), $F_{1, 1379}=61.3$, $p<0.001$,$\eta_p^2 = .04$. A more dramatic reduction was observed in rotational head velocity. In the Natural condition, the mean rotational speed was M = $133.30$ degree/s, SD = 88.13 degree/s. In the Passthrough condition, this dropped substantially to M = $33.41$ degree/s, SD = 11.70 degree/s: a four-fold decrease in the speed of head rotation, $F_{1, 1379}=374$, $p<0.001$, $\eta_p^2 = .21$.

Compared to normative values found in the literature, the Natural setting of this study falls within the range for an active task (90 - 180 degrees/s) (Grossman et al. \cite{grossman1988frequency}). However, the Passthrough setting is well below the threshold for ``natural active behavior.'' However, the head translational velocity for both settings are within the lower bound of typical values ($0.15 - 1.80$ m/s) (Pozzo et al. \cite{pozzo1990head}).

\begin{figure}[h!t]
\centering
\begin{minipage}[t]{.5\linewidth}
  \centering
  \captionsetup{margin=5pt}
  \includegraphics[height=5cm]{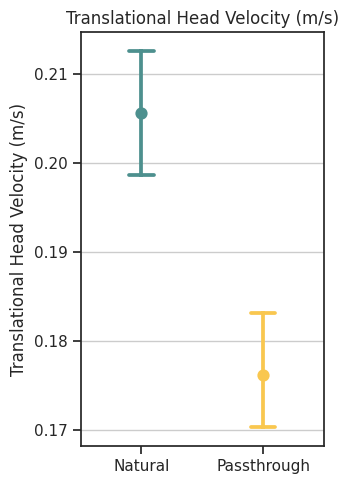}
  \captionof{figure}{Average Translational Head Velocity measured in m/s.}
  \label{fig:head_move_freq}
\end{minipage}%
\begin{minipage}[t]{.5\linewidth}
  \centering
  \captionsetup{margin=5pt}
  \includegraphics[height=5cm]{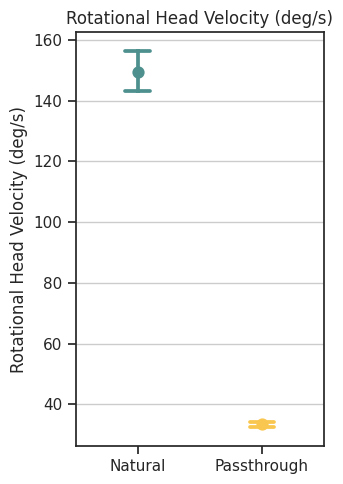}
  \captionof{figure}{Average Rotational Head Velocity measured in degree/s.}
  \label{fig:head_orien_freq}
\end{minipage}
\end{figure}

Figure \ref{fig:head_move_freq} and \ref{fig:head_orien_freq} illustrates these differences. The translational speed (Figure \ref{fig:head_move_freq}) and the rotational speed (Figure \ref{fig:head_orien_freq}) show a stark contrast for both Natural and Passthrough condition. In the Natural condition the rotational speed exhibits a broader distribution, indicating frequent rapid head turns. Conversely, the Passthrough distribution is extremely compressed near the baseline, indicating that participants almost entirely suppressed rapid head rotations.

\textbf{Hand Kinematics} Hand kinematics followed a similar, albeit subtler, pattern of reduction. Participants moved their hands statiscally significantly faster in the Natural condition (M = $0.32$ m/s, SD = 0.106 m/s) compared to the Passthrough condition (M = $0.31$ m/s, SD = 0.095 m/s), though the effect size was trivial, $F_{1, 1379}=7.94$, $p=0.005$, $\eta_p^2 = .01$. However, the speed of hand rotation was significantly higher in the Natural condition (M = $71.03$ degree/s, SD = 18.97 degree/s) compared to the Passthrough condition (M = $67.38$ degree/s, SD = 16.99 degree/s), $F_{1, 1379}=18.4$, $p<0.001$, $\eta_p^2 = .01$.

Translational hand kinematics for either viewing type are slightly below the normative range of $0.35 - 0.5$ m/s (Jeannerod et al. \cite{jeannerod1984timing}) and on the higher end for rotational hand kinematics ($30 - 70$ degree/s) (Fan et al. \cite{fan2021effects}).

\begin{figure}[h!t]
\centering
\begin{minipage}[t]{.5\linewidth}
  \centering
  \captionsetup{margin=5pt}
  \includegraphics[height=5cm]{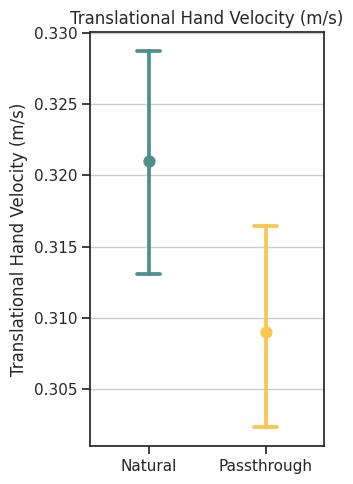}
  \captionof{figure}{Average hand movement measured in m/s.}
  \label{fig:hand_move_freq}
\end{minipage}%
\begin{minipage}[t]{.5\linewidth}
  \centering
  \captionsetup{margin=5pt}
  \includegraphics[height=5cm]{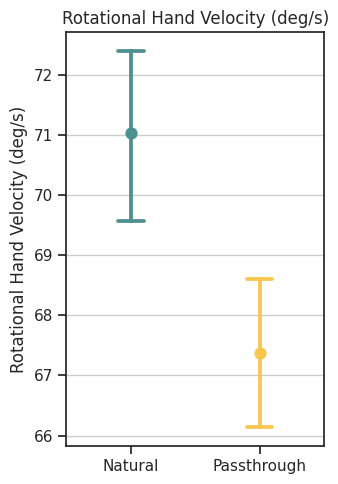}
  \captionof{figure}{Average hand rotation measured in degree/s.}
  \label{fig:hand_orien_freq}
\end{minipage}
\end{figure}

The analysis of hand kinematics revealed statistically significant differences between the Natural and Passthrough conditions. As illustrated in Figure \ref{fig:hand_move_freq}, the mean Hand Movement Velocity was higher in the Natural condition ($\approx 0.321$ m/s) compared to the Passthrough condition ($\approx 0.309$ m/s), though the practical effect size was trivial, $F(1, 1379) = 7.94, p = .005, \eta_p^2 = .01$. A similar trend was observed for Hand Rotation (Figure \ref{fig:hand_orien_freq}), where the Natural condition (M $\approx 71.0$ degree/s, SD = 18.9 degree/s) exceeded that of the Passthrough condition with a small effect size (M $\approx 67.4$ degree/s, SD = 16.9 degree/s), $F(1, 1379) = 18.40, p < .001, \eta_p^2 = .01$. Both results shows a consistent downward shift in medians, confirming a generalized ``motor dampening'' effect.

\textbf{Head Rotation And Fixations} To investigate the coupling between head movement and gaze, we analyzed the change in head rotation that occurred between two consecutive fixations. The results show that participants made on average significantly larger head movements between fixations in the Natural condition ($30.79^\circ$) compared to the Passthrough condition ($16.08^\circ$), $F_{1, 1379}=5.14$, $p<0.001$, $\eta_p^2 = .01$.

\begin{figure}[h!t]
\centering
        \includegraphics[width=1.0\linewidth]{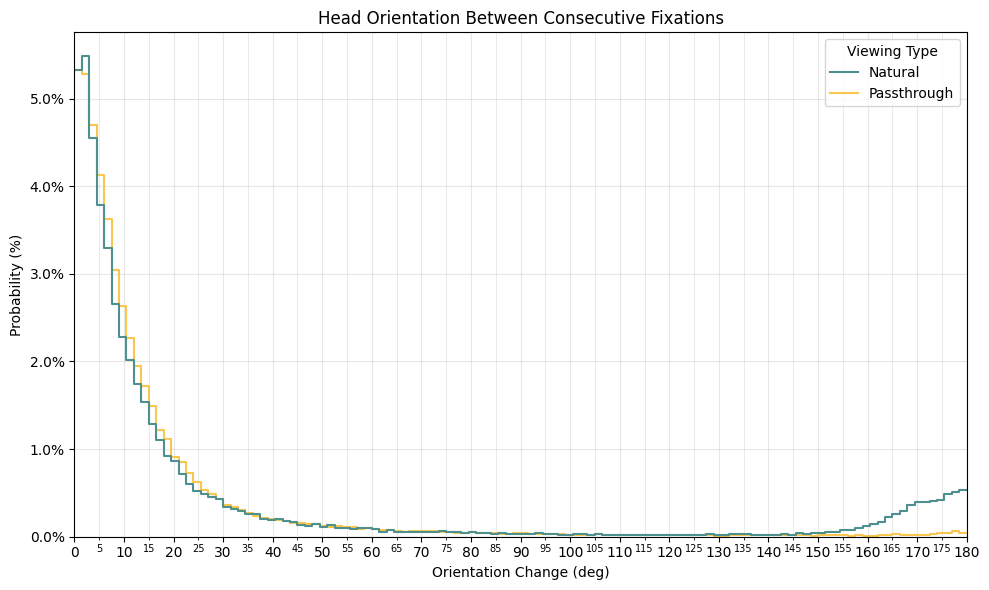}
    \caption{Orientation change of the head between two consecutive fixations measured in degrees.}
    \label{fig:head_move_fix}
\end{figure}

Figure \ref{fig:head_move_fix} presents the histogram of these orientation changes. The Natural condition (teal) displays a ``long tail'' with bimodal characteristics, extending toward $180^\circ$, indicating that participants occasionally made large head movements to shift their attention between fixations.  In contrast, the Passthrough condition (yellow) runs out mostly entirely past $110^\circ$, suggesting that gaze shifts in the mediated environment were primarily achieved through eye movements or very small head adjustments, rather than large, exploratory head rotations as seen in the natural setting.

\subsection{Analysis of Subjective Experience}

Finally, we evaluated the subjective physiological impact of the viewing conditions using the Simulator Sickness Questionnaire (SSQ). Participants rated their experience on a scale from ``None'' to ``Severe'' immediately following each trial. The analysis revealed significant degradation in comfort across ocular, cognitive, and general physical domains when using the passthrough system.

\textbf{Oculomotor Discomfort} The most pronounced effects were observed in variables related to visual stress. The viewing condition had a moderate, highly significant effect on Eye Strain ($F(1, 1561) = 75.40, p < .001, \eta_p^2 = .05$). The vast majority of participants reported ``None'' for Eye Strain in the Natural condition ($90.7\%$). In the Passthrough condition, this dropped to $69.5\%$, with a corresponding increase in ``Slight'' ($25.8\%$) and ``Moderate'' ($3.3\%$) reports.

%\begin{figure}[h!t]
%\centering
%        \includegraphics[width=1.0\textwidth]{img/selection/SSQ/eye_strain.jpg}
%    \caption{Eye Strain}
%    \label{fig:eye_strain}
%\end{figure}

This trend was mirrored in Difficulty Focusing ($F(1, 1561) = 74.00, p < .001, \eta_p^2 = .05$) and Blurred Vision ($F(1, 1561) = 50.50, p < .001, \eta_p^2 = .03$). %Figure \ref{fig:diff_focusing} illustrates the results for Difficulty Focusing. 
In the Natural condition, only $12.7\%$ of participants reported any difficulty. In contrast, under Passthrough viewing, $34.5\%$ of participants reported some degree of difficulty focusing, with ``Moderate'' reports rising from $3.2\%$ to $9.8\%$. Similarly, %Figure \ref{fig:blur_vision} shows that
reports of Blurred Vision shifted substantially, with the ``None'' category dropping from $91.8\%$ in Natural to $74.7\%$ in Passthrough.

%\begin{figure}[h!t]
%\centering
%        \includegraphics[width=1.0\textwidth]{img/selection/SSQ/dif_focusing.jpg}
%    \caption{Difficulty Focusing}
%    \label{fig:diff_focusing}
%\end{figure}

%\begin{figure}[h!t]
%\centering
%        \includegraphics[width=1.0\textwidth]{img/selection/SSQ/blurred_vision.jpg}
%    \caption{Blurred Vision}
%    \label{fig:blur_vision}
%\end{figure}

\textbf{Systemic Discomfort and Concentration} The cumulative effect of these visual stressors appeared to impact general well-being and cognitive load. General Discomfort showed a large, significant main effect of viewing condition ($F(1, 1561) = 168.00, p < .001, \eta_p^2 = .10$). %Figure \ref{fig:general_disc} highlights 
The data shows a substantial shift: while $76.8\%$ of participants reported no discomfort in the Natural condition, less than half ($43.7\%$) remained symptom-free in the Passthrough condition. Reports of ``Slight'' discomfort more than doubled (from $17.8\%$ to $41.6\%$), and ``Moderate'' discomfort increased nearly threefold (from $4.4\%$ to $12.0\%$).

%\begin{figure}[h!t]
%\centering
%        \includegraphics[width=1.0\textwidth]{img/selection/SSQ/general_discomfort.jpg}
%    \caption{General Discomfort}
%    \label{fig:general_disc}
%\end{figure}

Additionally, participants reported significantly greater Difficulty Concentrating in the Passthrough condition ($F(1, 1561) = 47.90, p < .001, \eta_p^2 = .03$). %As depicted in Figure \ref{fig:dif_conc}, 
The proportion of participants reporting ``None'' dropped from $82.1\%$ in Natural to $65.6\%$ in Passthrough, with a concurrent rise in ``Slight'' ($24.1\%$) and ``Moderate'' ($8.9\%$). This suggests that the physiological demands of the mediated environment competed with the cognitive resources available for the task.

%\begin{figure}[h!t]
%\centering
%        \includegraphics[width=1.0\textwidth]{img/selection/SSQ/general_discomfort.jpg}
%    \caption{General Discomfort}
%    \label{fig:dif_conc}
%\end{figure}

\section{Discussion}

The results of this study provide a quantitative depiction of the impact imposed by video passthrough mediation. This data suggests an interpretation that the reduced task efficiency observed in the Passthrough condition is not merely a result of visual degradation, but the outcome of a complex compensatory strategy where users trade biomechanical flexibility for sensory stability, incurring a significant physiological cost in the process.

\subsection{The ``Passthrough Rigidity'' Effect}

The most pronounced behavioral adaptation observed was the reduction in head rotation. While translational head movement showed only a minor decrease, rotational velocity dropped four-fold in the Passthrough condition. This behavior, an element of ``Passthrough Rigidity'' as defined earlier, likely serves as an adaptative strategy for motor caution. Rapid head rotations in passthrough often exacerbate motion-to-photon latency artifacts (Stauffert et al. \cite{stauffert2020latency} and Buker et al. \cite{buker2012effect}) and geometric distortions (El Chemaly et al. \cite{el2025mind}), which are known drivers of sensory conflict. By limiting these rotations, participants effectively minimized the vestibular-visual mismatch. Furthermore, this rigidity directly contrasts with the ``head-first'' scanning paradigm typically observed in restricted FOV conditions (Moss et al. \cite{moss2011characteristics}). Our interpretation is further supported by the decoupling of head and eye movements; in the Natural condition, gaze shifts were accompanied by significant head re-orientation ($30.79^{\circ}$), whereas in Passthrough, the head remained relatively static ($16.08^{\circ}$). These results are supported by the findings of Astrologo et al. \cite{astrologo2024determining} which find that the physical offset of the headset shifts the head's center of mass to the front, substantially increasing the moment of inertia around the neck (Astrologo et al. \cite{astrologo2024determining} and Willemsen et al. \cite{willemsen2004effects}). To avoid the increased torque required for rapid head rotations, users adopt a compensatory motor strategy, suppressing angular acceleration while relying on larger torso movements for translation. However, as our results show, this is not the only contributing factor.

\subsection{Visual Processing Efficiency}

With the head significantly more stabilized, the burden of information gathering shifted more to the oculomotor system, which showed clear signs of processing load. Although the total number of fixations remained comparable, the temporal structure of these fixations changed significantly. The increase in mean fixation duration suggests that participants required more time to perceive visual information from the digital feed. This aligns with the subjective reports of ``Blurred Vision'' and ``Difficulty Focusing'', suggesting that the limiting factor was the clarity of the visual input. This creates an ``Oculomotor Load Paradox'': while users rigidly stabilized their heads to avoid latency artifacts, they simultaneously exhibited extended fixation durations and a more centrally clustered spatial fixation distribution. Furthermore, the spatial distribution analysis showed that fixation distributions were slightly tighter and more centrally clustered in the Passthrough condition, and the Kullback–Leibler divergence analysis revealed a moderate, measurable difference. This implies that users not only restricted their visual area to a smaller window, possibly to avoid peripheral distortions or lens aberrations common in HMD optics but also they changed their eye movement behaviour and had to alter the way they use their eyes.

\subsection{Physiological and Cognitive Costs}

These behavioral adaptations allowed participants to complete the tasks, but they came at a measurable physiological price, besides lower accuracy and longer response times. The significant reduction in blink duration and the high prevalence of reported Eye Strain  point to a state of heightened ocular stress. Unlike the ``blink suppression'' often seen in high-cognitive-load tasks (where rate decreases Eckstein et al. \cite{eckstein2017beyond}), the shortening of blink duration observed here may reflect a specific irritation response as also show in digital eye strain metrics as shown in Kaur et al. \cite{kaur2022digital} or a struggle to maintain tear film stability without interrupting visual intake. Moreover, the generalized ``motor dampening'' observed in our translational and rotational hand kinematics provides a quantitative baseline for the dexterity deficits frequently reported in mediated environments (Joyner et al. \cite{joyner2021comparison}), suggesting that users unconsciously slow their movements to compensate for the sensorimotor recalibration imposed by the passthrough pipeline (Biocca et al. \cite{biocca1998virtual}).

Cumulatively, this visual and motor rigidity translated into a cognitive burden. The ``Difficulty Concentrating'' reported in the SSQ correlates with the slower Response Times and reduced Accuracy. This suggests that a portion of the user's cognitive capacity was diverted from the block assembly task to the maintenance of the passthrough interface itself. The absence of severe vestibular symptoms (e.g., Nausea/Vertigo were less severe than Ocular symptoms)  indicates that adaptive motor caution was largely effective at preventing motion sickness, but transferred the load to the visual and cognitive systems.

\subsection{Design Implications}

The behavioral and physiological constraints identified in this study highlight three critical design implications for next-generation XR interfaces. First, Gaze-Contingent Reprojection should be prioritized; because users exhibit ``Passthrough Rigidity'' and suppress gross head rotations, rendering architectures must dynamically reproject the scene based on eye-tracking to compensate for restricted physical exploration. Second, Texture Boosting for Eye Strain is necessary to counteract the significant ocular discomfort (e.g., difficulty focusing, blurred vision) isolated in our SSQ analysis. Dynamically enhancing contrast and texture resolution at the focal point may mitigate the optical fatigue induced by passthrough displays. Finally, developers should adopt Bio-Inspired UI Bounds. Given the four-fold decrease in rotational head velocity and restricted gross kinematics, spatial interfaces must be constrained to a narrower, forward-facing field of view that respects the encumbered biomechanics of HMD users, ensuring core interactive elements do not necessitate exhaustive physical rotation.

\subsection{Limitations and Confounds}

While this study quantifies significant behavioral shifts under passthrough viewing, several confounding variables must be prominently acknowledged. Foremost, the physical ergonomics of the HMD -- specifically its added weight and altered center of mass -- inherently restrict natural head kinematics independently of visual processing demands (Willemsen et al., \cite{willemsen2004effects}). Consequently, comparing an unencumbered Natural condition to an encumbered Passthrough condition convolutes these biomechanical constraints with actual display effects. Furthermore, the novelty of initial HMD use serves as a highly prominent confound in this study. Given that 51\% of our subjects reported no prior XR experience, the observed ``Passthrough Rigidity'' may be partially attributed to hardware unfamiliarity and cautious motor strategies rather than solely the optical limitations of the passthrough video.

\section{Conclusion and Future Work}

Using a novel methodology, this study quantified the multi-modal impact of a current-generation passthrough device. Our results demonstrate that while users can complete manual tasks successfully in a mediated environment, they do so by adopting a compensatory strategy that prefers sensory stability at the expense of biomechanical flexibility. Specifically, we observed a distinct effect, where participants suppressed rotational head velocity by a factor of four and decoupled head-gaze coordination to mitigate motion-to-photon latency artifacts and geometric distortions, contributing to ``Passthrough Rigidity.''

This kinematic restriction shifted the burden of information gathering to the oculomotor system. The significant increase in fixation duration, combined with a restricted spatial search pattern, indicates that resolving the visual feed required a higher degree of temporal processing effort than natural vision. While effective for maintaining accuracy, this adaptation incurred a significant physiological cost, evidenced by increased eye strain, difficulty concentrating, and general discomfort. Ultimately, passthrough forces users to operate in a state of heightened motor caution and cognitive load, creating a measurable trade-off between comfort and task performance.

Future work should investigate the longitudinal effects of the increased kinematic and oculomotor rigidity. Since our experimental sessions were limited to approximately one hour, it remains unclear whether the observed physiological costs plateau or compound during the extended use scenarios envisioned for spatial computing (e.g., an 8-hour workday). While this study isolated the effects of a single commercially available passthrough system, further research is required to determine if higher-resolution displays, lower-latency pipelines, light guards or a wider Field of View can effectively counter elements of ``Passthrough Rigidity,'' or if this is a fundamental human response to mediated perception. Additionally, future experiments should include other passthrough devices (such as Samsung Galaxy XR and Apple Vision Pro) and Optical See-Through (OST) devices (such as XReal Aura) to determine if kinematic rigidity is a universal symptom of mediated vision. Finally, this methodology can inform designers of future XR devices, providing a holistic signal of user comfort to guide hardware and software optimizations, including real-time predictive models as a function of these kinematic and oculomotor markers.

%% if specified like this the section will be omitted in review mode
\acknowledgments{%
  This work was supported in part by grants from Natural Sciences and Engineering Research Council of Canada (Grant Number RGPIN-2022-04606) Google Canada Corp.
}

\bibliographystyle{abbrv-doi-hyperref}

\bibliography{template}

\end{document}